\documentclass[prl,aps,twocolumn,nofootinbib]{revtex4}
\usepackage{graphicx}
\usepackage{hyperref}
\usepackage{amsmath,amssymb}
\usepackage{xcolor}
\usepackage{physics}

\definecolor{darkBlue}{rgb}{0, 0, 0.8}

\hypersetup{
  bookmarksopen=true,     
  colorlinks=true,        
  linkcolor=darkBlue,     
  citecolor=darkBlue,      
  filecolor=darkBlue,      
  urlcolor=darkBlue        
}

\def\be{\begin{equation}}
\def\ee{\end{equation}}
\def\ba{\begin{align}}
\def\ea{\end{align}}

\def\lsim{\raise0.3ex\hbox{$\;<$\kern-0.75em\raise-1.1ex\hbox{$\sim\;$}}}
\def\gsim{\raise0.3ex\hbox{$\;>$\kern-0.75em\raise-1.1ex\hbox{$\sim\;$}}}

\def\theta{\vartheta}

\begin{document}

\title{Comment on ``Dark Matter Annihilation Can Produce a Detectable Antihelium Flux through $\bar\Lambda_b$ Decays''}

\author{M.~Kachelrie\ss$^{1}$}
\author{S.~Ostapchenko$^{2}$}
\author{J. Tjemsland$^{1}$}

\affiliation{$^{1}$Institutt for fysikk, NTNU, Trondheim, Norway}

\affiliation{$^{2}$D.V. Skobeltsyn Institute of Nuclear Physics,
 Moscow State University, Moscow, Russia}

\maketitle

In a recent Letter, Winkler and Linden~\cite{WL} (hereafter WL21) suggested
that a previously neglected standard model process, namely the production of
antihelium-3 nuclei through decays of $\bar\Lambda_b$ baryons, can
significantly boost the flux of antihelium-3, induced by annihilations or decays
of dark matter. This suggestion uses the fact that  dark matter particles will
annihilate typically into the heaviest quark--anti-quark pair, i.e.\ $\bar b b$
pairs, if the particle is a Majorana fermion and its mass is below the mass
of the standard model gauge bosons~\cite{GB}. These (anti-) $b$ quarks will
in turn hadronise and form (anti-) $b$-mesons and (anti-) $b$-baryons which
then decay weakly. As pointed out by WL21, the $\bar\Lambda_b$ baryon is
especially suited for the production of antihelium-3 through a coalescence
process, because its rest mass of 5.6\,GeV is not much above the rest mass
of 5~(anti)-nucleons. As a result of the small relative momenta of these nucleons,
the production of antihelium-3 via coalescence  is enhanced
in  $\bar\Lambda_b$ decays.

A condition that the scenario of WL21 leads to a detectable
antihelium flux is that the branching ratio ${\rm BR}(b\to\Lambda_b)$ is
sufficiently large. In order to achieve such a large  branching ratio,
WL21 increased the diquark formation parameter {\tt probQQtoQ} of Pythia
in their so-called ``$\bar\Lambda_b$ tune''. WL21 noted that this change
also significantly boosts prompt antinucleon production.  They compensated
the resulting over-production of antideuterons by reducing at the same time
the coalescence momentum, which is  a free parameter in their approach, by
a factor~0.6.

\begin{table}[!b]
\begin{center}
\begin{tabular}{c|c|c|c|c}
  $\sqrt{s}$  & $\approx10$\,GeV & 29--35\,GeV & 91\,GeV & 130--200\,GeV \\
  Obs.  & $0.266\pm 0.008$ & $0.640\pm0.050$  & $1.050\pm 0.032$ & $1.41\pm0.18$\\
  WL21  & 0.640 & 1.161  & 2.102 & 2.33 \\ 
\end{tabular}
\end{center}
\caption{Multiplicity of (anti-) protons in electron-positron
annihilations}
%
\begin{center}
\begin{tabular}{c|c|c|c}
  Particle  & proton & kaon & pion \\
  $\dv*{N}{y}$, LHC & $0.124\pm0.009$ & $ 0.286\pm0.016$ & $2.26\pm 0.10$ \\
  $\dv*{N}{y}$, $\Lambda_b$ tune & 0.328 & 0.231 & 1.90 \\ 
\end{tabular}
\end{center}
\caption{Measurements of $\dv*{N}{y}$ at mid-rapidity ($|y|<0.5$) in proton-proton
collisions at $\sqrt{s}=7$\,TeV for $p$, $K^+$ and $\pi^+$} 
\end{table}

The conceptual error of WL21 is that the change of {\tt probQQtoQ} cannot
simply be compensated by a reduction of the coalescence momentum, since this
change affects all types of processes involving baryon and meson production.
As an example, one can consider \mbox{(anti-)} proton production
in electron-positron annihilations, $e^+e^-\to \bar p pX$. For a change
of {\tt probQQtoQ} from the default value 0.09
to 0.24---which is the value reproducing the value of the branching ratio
$b\to\Lambda_b=0.1$ chosen in WL21---the resulting proton multiplicity is
compared in Table~1 to measurements. For instance at
$\sqrt{s}=91$\,GeV,
the predicted proton multiplicity in the ``$\bar\Lambda_b$ tune'' is $33\sigma$
away from the one measured~\cite{PDG}.
For comparison, the standard settings in Pythia predict a $\Lambda_b$ multiplicity in
electron-positron annihilations at the $Z$-resonance of $0.016$, which is less than
$1\,\sigma$ away from the value $0.031\pm0.016$ given in Ref.~\cite{PDG}.
As an example  for the effects of a changed diquark formation parameter
on $pp$ collisions, we show in Table~2 the integrated 
yield at mid-rapidity, $\dv*{N}{y}|_{|y|<0.5}$, of protons, kaons and pions 
measured by ALICE at LHC at $\sqrt{s}=7$\,TeV \cite{ALICE}. Note also  that the
increased diquark formation reduces the production rate of all mesons,
aggravating the variance of the ``$\bar\Lambda_b$ tune'' with observations.
Finally, the condition not to overproduce the antiproton
flux  measurements~\cite{AMS-02} from AMS-02 requires to reduce the
annihilation rate of dark matter in the ``$\bar\Lambda_b$ tune''
relative to the value allowed using the default version of Pythia.

Another caveat in the approach of WL21 is the use of Pythia to ``predict''
the branching ratio ${\rm BR(\bar\Lambda_b\to \bar u d u (ud_0))}=0.012$
which controls the formation rate of antihelium-3.  Such ratios are external
input parameters into Pythia, which represent in the case of yet unobserved
decays simply educated guesses.
In this specific case, the ratio is $\simeq |V_{ub}|^2/|V_{cb}|^2$, while one
expects an additional suppression if diquarks are formed.
Comparing branching ratios of such $\bar\Lambda_b$ decays
to observations, we find indeed that Pythia overestimates their rate by a
factor 4--5, which is further enhanced in the $\Lambda_b$ tune. In particular,
Pythia using the standard settings overestimates the measured branching ratio
${\rm BR(\bar\Lambda_b\to \Lambda_c^-\bar pp\pi^+)}=(2.65\pm0.29)\times 10^{-4}$ \cite{PDG}
by a factor 5.6 and in the $\Lambda_b$-tune by a factor 17. This corresponds to 
a $42\sigma$ and $144\sigma$ deviation from measurements, respectively.
Reducing  ${\rm BR(\bar\Lambda_b\to \bar u d u (ud_0))}$ correspondingly
would make the antihelium-3 flux undetectable for AMS-02 even if the
``$\bar\Lambda_b$ tune'' would be viable.

In conclusion, the ``$\bar\Lambda_b$ tune'' of Pythia which WL21 argue
to lead to an antihelium-3 flux detectable by AMS-02 is excluded by a wealth
of measurements of (anti-) baryon and (anti-) meson production at accelerators.
Even so, the future observation of  (anti-) helium-3 production in baryon
decays can potentially have a profound impact on the study of hadronisation,
as noted already by WL21.
This rate varies by orders of magnitude between  event generators based on
different hadronisation models and may thus be used to discriminate these
models.

\acknowledgments
We would like to thank Philip Ilten, Uli Nierste and Torbj{\"o}rn
Sj{\"o}strand for helpful comments.

\end{document}